# The role of local field modes in calculating the effective optical properties of metamaterials

Anderson O. Silva,[1,2] Ivo T. Leite,[1,3] José M. Teixeira,[3,4] João C. W. A. Costa,[2] Maria T. M. R. Giraldi,[5] Ariel Guerreiro[1,3]

[1]INESCTEC-UOSE (National Institute of System Engineering and Computers – Unit of Optoelectronics and Electronic Systems), R. Campo Alegre – 4169-007, Porto, Portugal.

[2]LEA-UFPA (Federal University of Para – Laboratory of Applied Electromagnetism), C. U. José da Silveira Neto – 66075-110, Belem, Brazil.

[3]Faculty of Sciences – University of Porto, R. Campo Alegre – 4169-007, Porto, Portugal.

[4]IFIMUP (Institute of Material Physics – University of Porto), R. Campo Alegre – 4169-007, Porto, Portugal

[5]IME – Military Institute of Engineering, Urca - 22290-270, Rio de Janeiro, Brazil.

This letter discusses the importance of the nature of local fields inside a unit cell of optical metamaterials in developing a valid effective medium theory. In particular, we apply the Meier-Wokaun procedure to derive a corrected version of the Maxwell-Garnett formula for the refractive index of a medium composed by metallic rods with subwavelength cross-section. We compare this result with the values obtained using the full mode analysis calculated via numerical simulations and analyze the impact of surface plasmon polaritons in adequately determining the effective refractive index of the medium.

The concept of metamaterials refers to man-made structures carefully engineered and built to display an exotic response to electromagnetic radiation which cannot be found in naturally occurring materials.[1] These response properties result from subwavelength structures (designated as meta-atoms) of different materials displayed in a regular pattern.

Historically, the first metamaterials operated in the microwave range.[2,3] However, advances in nanofabrication techniques have allowed to reduce the size of the unit cell to nanoscopic dimensions and extended the operation wavelength to the optical domain.

Among the later are included periodic arrays of elongated metallic cylinders (designated as nanowires) immersed in a dielectric medium. The nonconventional optical properties of these media have been verified experimentally and include negative refraction,[4] hyperlens imaging,[5] local field enhancement,[6] hyperbolic dispersion,[7] to name a few. Like in most optical metamaterials, the special electromagnetic properties of nanowires result mainly from the excitation of surface plasmon polariton modes at the metal-dielectric interface, which allow to guide light below the diffraction limit. The theoretical formalism in addressing the propagation of optical beams in metallic cylinders with nanometer diameter was summarized by Takahara *et al.*.[8] Further contribution to this theory can be found in Ref. 9-10, in which the propagation of plasmon modes bounded to the nanowire surface is characterized in terms of losses and in accordance to the surrounding dielectric.

The basic idea behind most of the models of metamaterials is that the interaction of light with subwavelength structures can be equivalently described in terms of macroscopic effective parameters corresponding to a homogeneous optical medium. The differences between several homogenization models of metamaterials depend mainly on two choices: i) how to compute the macroscopic fields and parameters from the local fields induced by an incident electromagnetic wave in the structure of the meta-atom, and ii) how to estimate these local fields. Typically, the first choice is made between the Bruggeman and the Maxwell-Garnett (MG) approaches,[11] which are extensions of Clausius-Mossotti formula. All these approaches determine the macroscopic fields by considering the reaction fields induced in an inclusion embedded in a host medium but they differ in what is considered inclusion and host.

In the Bruggeman approach, both the dielectric and metallic structures are considered inclusions while the host medium is the homogenized material itself. This is adequate to describe structures that have a low refractive index contrast between the materials that composes the meta-atom, which is not the case of metal-dielectric interfaces. Then, the validity of this model is limited to randomly distributed inclusions in the long wavelength approximation. In the MG approach, the metal is considered to be the inclusion while the dielectric is the host. Unlike the Bruggeman theory, this model can adequately describe metamaterials with high refractive index contrast and is usually preferred in the literature to model metallic nanowires in a dielectric background.[12,13]

Regarding the second choice, specifically dealing with estimation of the local fields, the original MG theory computes the macroscopic parameters basically from the average of the electrostatic fields generated by the dipoles induced in spherical inclusions. This is sometimes referred as the quasi-static approximation and corresponds to the first term of the expansion of the scattered fields obtained in the Mie theory.[14] Therefore, it totally neglects the real structure of the meta-atom. To overcome this limitation, several corrections have been proposed to better estimate the fields induced in the meta-atoms by the incident electromagnetic wave and which strongly depend on their geometry and material composition. These corrections can be expressed in the effective permittivity in terms of a depolarization factor $\zeta_{eff}$, which in general is a tensor.[15] A qualitative discussion on the correction introduced by $\zeta_{eff}$ in the Bruggeman and MG models was carried out in Ref. 16 using the assumption of a quasi-static approximation. It addresses the case where all the dimensions of the inclusions are in subwavelength size. To extend beyond this, Meier and Wokaun[17] derived a model that includes the field enhancement on large spherical metallic particles to obtain a dynamical counterpart for $\zeta_{eff}$. This dynamical term corrects the polarization $\overline{P}$ for the spatial dephasing between each infinitesimal dipole moment induced within the particle volume. From these calculations arises an additional imaginary contribution, proportional to the total particle volume and is attributed to radiative losses. Foss *et al.*[18] substituted the static term of the Meier-Wokaun model for spheres by the corresponding value for thin circular cylinders to retrieve the effective permittivity of an arrangement of nanoscopic needle-shaped gold particles. Liu *et al.*[19] used this modified expression as an extension of the MG model to compute the effective permittivity of an hyperbolic-dispersion metamaterial composed by silver nanowires. Moroz[20] extended the Meier-Wokaun approach to prolate and oblate spheroidal particles in the cases when the applied electric field is both perpendicular and parallel to the principal axis.

In this paper, we revisit the Meier-Wokaun approach to formulate rigorously the dynamical MG model for cylindrical structures. To assess the accuracy of this model, we use it to calculate the effective refractive index of an optical metamaterial composed by a periodic array of silver nanowires immersed in an alumina background and compare it with the refractive indexes obtained from numerical solutions of Helmholtz equation.

According to the MG model, the effective dielectric constant $\varepsilon_{eff}$ of well-ordered inclusions immersed in a host matrix is:[19]

$$\varepsilon_{eff} = \varepsilon_d + \frac{f\varepsilon_d(\varepsilon_m - \varepsilon_d)}{\varepsilon_d + \zeta_{eff}(1-f)(\varepsilon_m - \varepsilon_d)}, \tag{1}$$

where $\varepsilon_m$ and $\varepsilon_d$ are the dielectric constants for the metallic inclusions and the dielectric background, respectively, $f$ is the filling-ratio and $\zeta_{eff}$ is the depolarization factor.

As previously mentioned, a better estimative of $\zeta_{eff}$ comprises the static depolarization factor designated by $L$, as well as the dynamical and radiative components. Using the procedure proposed by Meier and Wokaun, we calculate analytically the depolarization factor $\zeta_{eff}$ for a cylindrical metallic structure with circular cross-section. We consider a finite cylinder with a general aspect ratio ($h/a$) between the height $h$ and the radius *a* and extrapolate the result for a nanowire by taking the asymptotic condition $h \gg a$.

Fig. 1 shows a schematic view of a silver nanowire medium. In this geometry, the nanowires are arranged in a hexagonal lattice immersed in an alumina background. The filling-ratio, the nanowire radius and the distance *D* between adjacent inclusions are related by $f = \left(\pi/\sqrt{12}\right)(2r/D)^2$. Following the Meier-Wokaun approach, we define an elemental depolarization electric field $d\overline{E}_d$ generated by an retarded dipole moment $d\overline{p} = e^{j\kappa r}\overline{P}dV$ in an infinitesimal volume $dV$ ($\kappa$ is the wavenumber, $r$ is the spherical radial coordinate and $\overline{P}$ is the polarization vector).[21] Expanding the resulting expression for $d\overline{E}_d$ in powers of $\kappa$ and considering only terms up to $\kappa^3$, we obtain:

$$d\overline{E}_d = \left\{ \frac{3\hat{r}(\overline{P}\cdot\hat{r}) - \overline{P}}{r^3} + \frac{\hat{r}(\overline{P}\cdot\hat{r}) + \overline{P}}{2r}\kappa^2 + j\frac{2}{3}\kappa^3\overline{P} \right\} dV . \tag{2}$$

We assume that $\overline{P}$ is uniform inside the metal cylinder and integrate $d\overline{E}_d$ over the volume to obtain:

$$\overline{E}_d = -4\pi\left( \overline{\overline{L}} - \frac{\kappa^2(2/3)V_C}{4\pi}\overline{\overline{C}} - j\frac{2\kappa^3}{4\pi}(2/3)V_C\overline{\overline{I}} \right) \cdot \overline{P} . \tag{3}$$

In (3), $V_C$ is the volume of the cylinder and $\overline{\overline{I}}$ is the unitary diagonal tensor. In our calculations, we consider that a plane wave with the electric field $\overline{E}_0$ lying on the azimuthal *xy* plane incides perpendicularly to the cross-section of the cylinder. In this case, the dynamical depolarization tensor $\overline{\overline{C}}$ is null and the transversal electrostatic depolarization tensor $\overline{\overline{L}}$ is:[15]

$$\overline{\overline{L}} = \begin{pmatrix} (1/2)\cos\theta & 0 \\ 0 & (1/2)\cos\theta \end{pmatrix}. \tag{4}$$

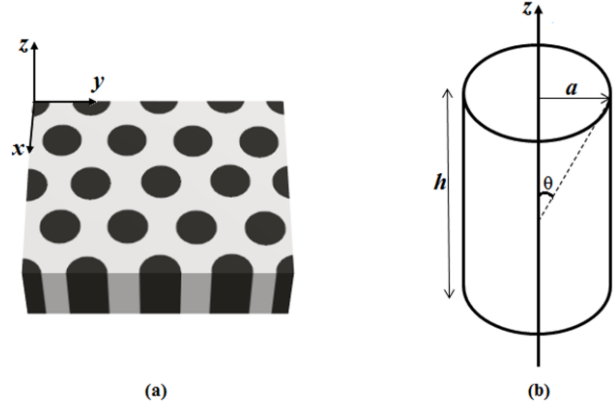

FIG. 1. (a) Scheme of the structure of the metamaterial composed by silver nanowires immersed in alumina background. (b) Graphical representation of the nanowire with indication of the relevant parameters.

The term $-4\pi\overline{\overline{L}}\cdot\overline{P}$ corresponds to the static contribution to $\varepsilon_{eff}$ whose derivation involves the regularization of the integral at the origin ($r=0$), as it is extensively discussed in Ref. 15. The contribution from the $\kappa^3$ term is straightforward to compute by simple integration. Applying (3) in the relation $4\pi\overline{P} = (\varepsilon_m - 1)(\overline{E}_0 + \overline{E}_d)$ and using the identity $V_c\overline{P} = \alpha_\perp \overline{E}_0$, the obtained expression for the polarizability $\alpha_\perp$ on the azimuthal plane is:

$$\alpha_\perp = \frac{V_C}{4\pi} \frac{\varepsilon_m - 1}{1 + \zeta_{eff}(\varepsilon_m - 1)}, \tag{5}$$

where the depolarization factor is:

$$\zeta_{eff} = \frac{1}{2}\cos\theta - j\frac{\kappa^3}{4\pi}\left(\frac{2}{3}\right)V_C. \tag{6}$$

We note that for cylindrical inclusions only the terms in $\kappa^0$ and $\kappa^3$ (respectively, the electrostatic and the damping terms) contributes to the correction of the MG model. This differs from the results obtained by Moroz for prolate spheroids where in addition there is a non-zero contribution from a term in $\kappa^2$, which is usually interpreted as a dephasing in the response of the infinitesimal dipoles located at different points of the meta-atom. Our results and the ones from Moroz only coincide in the asymptotic limit in which the major principal axis of the prolate spheroid becomes infinite and the shape of the inclusion is in practice an infinitely tall cylinder. Then, the contribution in $\kappa^2$ becomes null and the expression for the remaining terms converge to our results.

We apply (6) to (1) and calculate the effective refractive index of the nanowire medium depicted in Fig. 1(a). The dielectric constant of silver was fitted from the experimental results obtained by Johnson and Christy,[22] while it is assumed that the alumina background has a dielectric constant $\varepsilon_d=3$[23] at the optical range.

The accuracy of the corrected MG model is assessed by comparing the values of the refractive index with those predicted by solving numerically the Helmholtz equation:

$$\nabla^2 \overline{E} + \kappa_0^2 \left(n^2 - n_{eff}^2\right)\overline{E} = 0, \tag{7}$$

subjected to:

$$\overline{E}(x,y,,z,t) = A(x,y)\exp\left[j(\omega t - n_{eff}\kappa_0 z)\right], \tag{8}$$

where $n(x,y,z) = \sqrt{\varepsilon(x,y,z)}$ is the refractive index of the material in each point (*x,y,z*) of the meta-atom and $n_{eff}$ is the effective refractive index. *A(x,y)* is the field amplitude in the $\varepsilon(x,y)$ region and $\kappa_0$ is the wavenumber in free-space.

We solve (7) by using a FEM (Finite Element Method) algorithm.[24] The parameter $n_{eff}$ is computed by the numerical method as the generalized eigenvalue associated to each eigenmode. Therefore, it can be regarded as the result of a homogenization procedure without any of the approximations used in the MG model other than considering that the dielectric constant at each point of the structure coincides with the bulk values of the corresponding material. Notice that these eigenmodes includes surface plasmon polaritons.

The analytical and numerical results are plotted in Fig. 2 for two filling-ratios: *f*=10% and 17%. In both cases, the distance between adjacent nanowires is fixed in *D*=60 nm. The Bruggeman theory underestimates the resonance profile of the nanowire medium. Then, the effective refractive index obtained presents lower losses over a broader spectral range than the MG model. This results from neglecting the surface plasmon modes. Since this formulation does not consider a clear distinction between inclusions and background, the modes strongly localized at the interface between metal and dielectric cannot be adequately estimated and only bulk modes are taken into account. On the other hand, MG model distinguishes the contribution of the metallic inclusions and of the dielectric embedding matrix for the confined and radiated fields. The correction introduced by $\zeta_{eff}$ is responsible for capturing the geometry dependence of the plasmons at the silver-alumina interface and its collective effect on the optical properties of the material.

The effective refractive index calculated by the corrected MG model meets the bandwidth obtained from the eigenmode analysis. The sharp peak of the imaginary part of $n_{eff}$ is related to the high amount of electromagnetic energy that penetrates in the metal at the plasmon resonance. Furthermore, the resonance band increases and is shifted toward large wavelengths by increasing the filling-factor $f$.

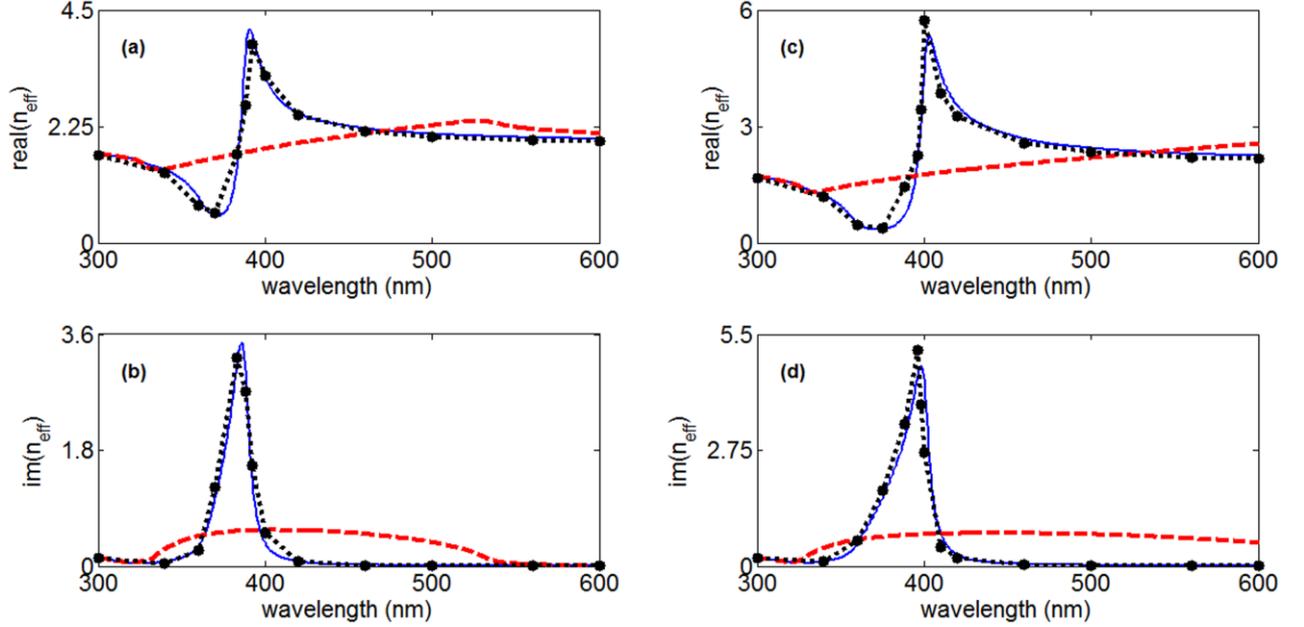

FIG. 2. (Color online) Real and imaginary parts of the effective refractive index for the lattice of silver nanowires immersed in an alumina background calculated from: Bruggeman theory (dashed line), corrected MG model (continuous line) and numerical eigenvalue solution (dot-circle line). Two cases are plotted: (a)-(b) $f$=10% and (c)-(d) $f$=17%.

The resonant wavelengths obtained from our analytical correction to the MG model are slightly red-shifted relatively to the ones calculated using the numerical eigenmode solutions. This is the consequence of the approximations assumed in deducing (6). Firstly, the depolarization field is computed by truncating the expansion $\overline{E}_d$ up to terms in $\kappa^3$, which is not required when using the numerical calculation. Secondly, the polarization induced is not homogeneous. The electromagnetic fields associated to surface plasmon modes are deeply attached to the metal-dielectric interface at the resonance, which can yield a polarization magnitude with strong variations along the cross-section radius. As shown in Fig. 3, an interesting aspect is that near the resonant wavelength the spatial structure of the plasmon mode changes dramatically at the interface between the metamaterial and air (the medium outside the metamaterial). Notice that there are two plasmon modes: one on the top surface of the nanowire and another along the cylindrical surface. Near the resonance, the coupling between these two plasmon modes is maximum and, as consequence, increases the field distribution.

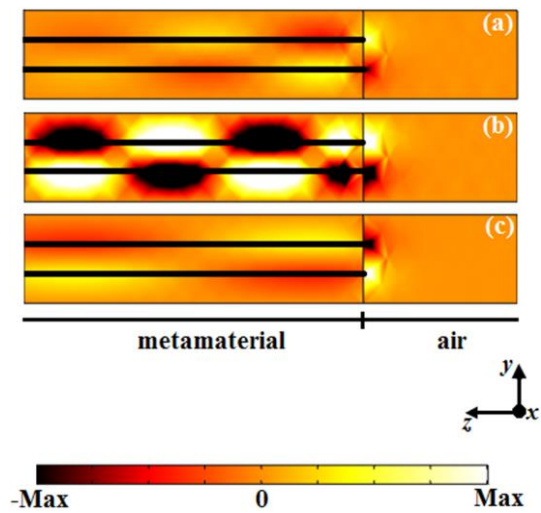

FIG. 3. (Color online) Electric field amplitude for three wavelengths along the spectrum of the nanowire medium with $f$=10%: (a) 300 nm, (b) 380 nm and (c) 600 nm. The silver nanowire is delimited by the thick black lines and the incidence is supposed normal to the horizontal interface. The resonance occurs at the wavelength $\lambda$=380 nm.

The results demonstrate the importance of considering the eigenmode structure of the field distribution along the meta-atom when calculating the effective optical properties of metamaterials. For metallic nanowires, in the spectral domain near the plasmon resonance, the structure of the plasmon modes undergoes a significant change, which account for the discrepancies between the exact numerical model and the approximated correction of the MG model presented here. Nanowires are relatively simple structures, however, for more complex meta-atoms it is expected that the localized electromagnetic modes should play an even important role in defining the optical properties of the medium.

**ACKNOWLEDGMENTS**

We thank A. Moroz for helpful discussion. This paper is within the framework of projects 275/2010, financially supported by CAPES (Brazilian Governmental Research Agency), and PG03109, financially supported by FCT (Portuguese Governmental Research Agency). The Brazilian authors would like to thank CNPq (Brazilian Governmental Research Agency) for partially also support this work. J. M. acknowledges support from FCT grant SFRH/BPD/72329/2010.